\magnification\magstephalf
\overfullrule 0pt
\input epsf.tex

\font\rfont=cmr10 at 10 true pt
\def\ref#1{$^{\hbox{\rfont {[#1]}}}$}


\font\fourteenbf=cmbx12 scaled\magstep1


\def\pd {\partial}
\def\pmb#1{\setbox0=\hbox{#1}
 \kern.05em\copy0\kern-\wd0 \kern-.025em\raise.0433em\box0 }

\def\slash{/\kern-.5em}

\def \half {{\scriptstyle {1 \over 2}}}

 %


\def\boxit#1{\vbox{\hrule\hbox{\vrule\kern1pt\vbox
{\kern1pt#1\kern1pt}\kern1pt\vrule}\hrule}}

\parskip=6pt
\parindent=0pt
\hsize=17truecm\hoffset=-5truemm
\vsize=23truecm
\def\footnoterule{\kern-3pt
\hrule width 17truecm \kern 2.6pt}


\catcode`\@=11 

\def\nolabels{\def\wrlabeL##1{}\def\eqlabeL##1{}\def\reflabeL##1{}}
\def\writelabels{\def\wrlabeL##1{\leavevmode\vadjust{\rlap{\smash%
{\line{{\escapechar=` \hfill\rlap{\sevenrm\hskip.03in\string##1}}}}}}}%
\def\eqlabeL##1{{\escapechar-1\rlap{\sevenrm\hskip.05in\string##1}}}%
\def\reflabeL##1{\noexpand\llap{\noexpand\sevenrm\string\string\string##1}}}
\nolabels
\global\newcount\refno \global\refno=1
\newwrite\rfile
\def\defref{$^{{\hbox{\rfont [\the\refno]}}}$\nref}
\def\defreflow{\the\refno\nref}
\def\nref#1{\xdef#1{\the\refno}\writedef{#1\leftbracket#1}%
\ifnum\refno=1\immediate\openout\rfile=refs.tmp\fi
\global\advance\refno by1\chardef\wfile=\rfile\immediate
\write\rfile{\noexpand\item{#1\ }\reflabeL{#1\hskip.31in}\pctsign}\findarg}
\def\findarg#1#{\begingroup\obeylines\newlinechar=`\^^M\pass@rg}
{\obeylines\gdef\pass@rg#1{\writ@line\relax #1^^M\hbox{}^^M}%
\gdef\writ@line#1^^M{\expandafter\toks0\expandafter{\striprel@x #1}%
\edef\next{\the\toks0}\ifx\next\em@rk\let\next=\endgroup\else\ifx\next\empty%
\else\immediate\write\wfile{\the\toks0}\fi\let\next=\writ@line\fi\next\relax}}
\def\striprel@x#1{} \def\em@rk{\hbox{}} 
\def\lref{\begingroup\obeylines\lr@f}
\def\lr@f#1#2{\gdef#1{\defref#1{#2}}\endgroup\unskip}
\newwrite\lfile
{\escapechar-1\xdef\pctsign{\string\%}\xdef\leftbracket{\string\{}
\xdef\rightbracket{\string\}}}

\def\writestop{\def\writestoppt{\immediate\write\lfile{\string\p
ageno%
\the\pageno\string\startrefs\leftbracket\the\refno\rightbracket%
\string\def\string\secsym\leftbracket\secsym\rightbracket%
\string\secno\the\secno\string\meqno\the\meqno}\immediate\closeout\lfile}}
\def\writestoppt{}\def\writedef#1{}
\catcode`\@=12 
\rightline{DAMTP 96/85}
\rightline{TUW 96-24}
\vskip 4mm
\centerline{\fourteenbf ELIMINATING INFRARED DIVERGENCES}
\vskip 2mm
\centerline{\fourteenbf IN THE PRESSURE} 
\vskip 6mm
\centerline{\bf I T Drummond, R R Horgan, P V Landshoff}
\vskip 1mm
\centerline{DAMTP, University of Cambridge$^*$}
\vskip 4mm
\centerline{\it and}
\vskip 4mm
\centerline{\bf A Rebhan}
\vskip 1mm
\centerline{Institut f\"ur Theoretische Physik, Technische Universit\"at Wien, 
Vienna$^*$}
\footnote{}{$^*$ itd@damtp.cam.ac.uk \  rrh@damtp.cam.ac.uk \ 
pvl@damtp.cam.ac.uk \ rebhana@tph16.tuwien.ac.at}
\vskip 0.5 cm
{\parindent=2cm\narrower\noindent{\bf Abstract}\par\noindent
The pressure of a system in thermal equilibrium is expressed as a mass integral
over a sum of thermal propagators. This allows a Dyson resummation and
is used to demonstrate that potential infrared divergences are rendered 
harmless.\par}

\vskip 0.5 cm

The perturbative treatment of finite-temperature quantum field theories  
involving massless bosonic fields always faces infrared divergences starting at
a certain loop order. In the case of the partition function or pressure,
these set in at three-loop order in four-dimensional 
theories. One relatively harmless
class of these infrared divergences has to do with the
appearance of thermal masses. Repeated self-energy insertions in a closed
loop give rise to arbitrarily high powers of products of these
thermal masses and massless propagators all with the
same momentum. However, once these thermal masses are resummed into
the propagators, 
such infrared divergences are avoided. As a remnant,
the resummed perturbative series is found to be nonanalytic in the conventional
loop expansion parameter --- instead of being organized in, say, powers of
coupling constant squared, the series then involves single powers and
logarithms of the coupling.

On the other hand, if there are modes for which
no thermal masses are generated, there is a different
source of infrared divergences at and beyond four-loop order
in renormalizable theories. This is
the case in
nonabelian gauge theories at finite temperature, 
where
perturbation theory fails to produce screening masses for
static chromomagnetic fields. Since in nonabelian gauge theories
there are vertices connecting exclusively magnetostatic modes, 
it is generally
assumed\defref\Linde{A D Linde, Phys Lett 96B (1980) 289;
D Gross, R Pisarski, L Yaffe, Rev Mod Phys 53 (1981) 43}\defref\lebellac{
M LeBellac, {\it Thermal field theory}, Cambridge University Press (1996)
} 
that perturbation theory has to stop at three
loop order. 
Perturbation theory seems to be capable only of establishing
the matching to the nonperturbative sector, which is that of
an effective lower-dimensional gauge theory and which has to be
treated by nonperturbative  methods, for example on a 
lattice\defref\Br{E Braaten,
Phys Rev Lett 74 (1995) 2164}.

An open problem of nonabelian gauge theories at finite temperature
is therefore whether something like the generation of
a thermal screening mass is indeed the mechanism which removes the
infrared divergences of
perturbation theory. But in this note we show that, in the case of the
pressure, a screening mass is not essential to make the
infrared divergences harmless. (Within the ladder approximation, something
similar has been put forward as a conjecture in Ref. \defreflow\daf{A P 
de Almeida and J Frenkel, Phys Rev D47 (1993) 640}.)

The pressure at temperature $T$ is given by
$$
P={T\over V}\log Z
\eqno(1a)
$$
where
$$\hbox{
\centerline{
\epsfxsize=12.97truecm
\epsfbox{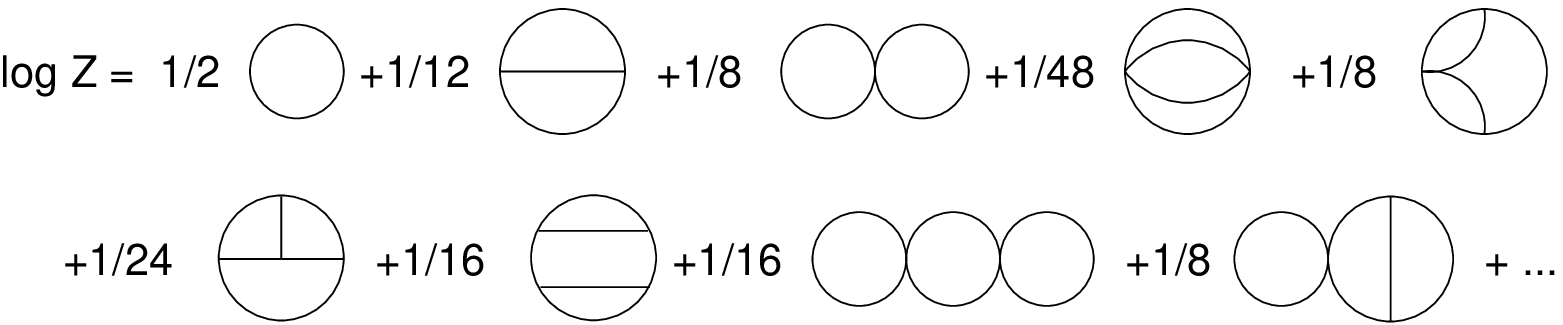} 
}
}\eqno(1b)$$
By naive powercounting\ref{\Linde}, all the
diagrams 
appear to be dangerous beyond three-loop order, and some of them already
at three-loop order (if there are unresummed thermal masses). 
Since they all can be looked at
as some subdiagram(s) inserted in one propagator closed onto itself,
one might expect that a Dyson resummation could take them into
account. But this expectation is spoiled by mismatching combinatorial
factors: making a single self-energy insertion in the first diagram of
log $Z$ gives
$$\hbox{
\centerline{
\epsfxsize=10.71truecm 
\epsfbox{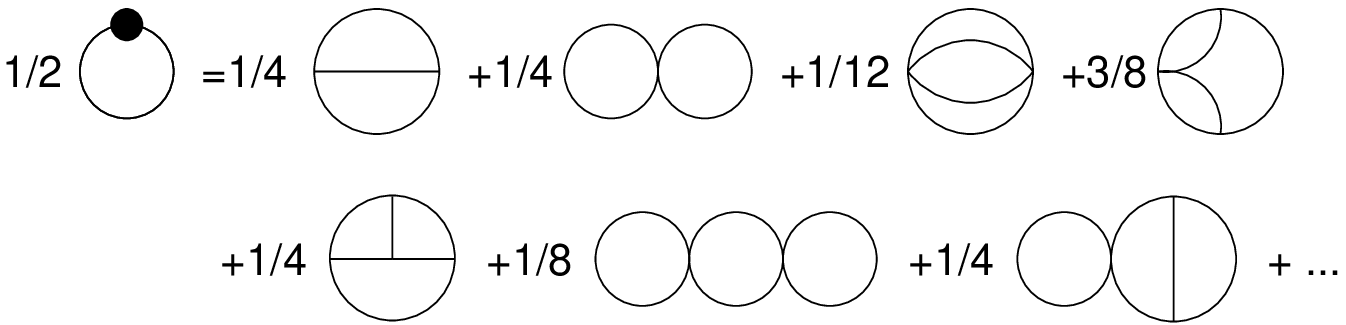}
}
}\eqno(1c)$$

In the traditional
ring-resummation\defref\Kapusta{J I Kapusta, {\it Finite-temperature field
theory}, Cambridge University Press (1989)}
method, one considers multiple self-energy insertions. 
This does give the correct
combinatorial factors, but only for those diagrams 
having at least two self-energy insertions, which is
sufficient for eliminating the infrared divergences caused by
unresummed thermal mass insertions. 
The case of 
only one irreducible self-energy 
insertion 
is not covered
and so we need some other method to see that there is no infrared problem
when there are propagators that are not sufficiently screened.

In the following, we shall derive an unconventional formula for the
partition function of a quantum field theory at finite temperature,
which has the form of a resummed one-loop quantity, while being
exact. 

The partition function $Z$ may be written\ref{\lebellac}
as a path integral:
$$
Z(T)=\int \prod _rd\phi _r \exp\left (iS[\phi ,T]\right )
\eqno(2)
$$
Here, the thermal action is written in terms of unrenormalised fields
and couplings:
$$
S[\phi ,T]=\int _C d^4x \left 
(\half \sum _r(\pd \phi _r.\pd\phi _r- m_r^2\phi _r^2)
   +{\cal L}_{{\sevenrm INT}}\right )
\eqno(3)
$$
In the variant of the real-time thermal field theory we are using,
the integration is over all {\bf x} and over the time-contour $C$ which,
in the complex $t$-plane, runs along the real axis from $-\infty$ to
$+\infty$, back to $-\infty$, and then down to $-\infty -i/T$.
The action has been written as if all the fields $\phi$ were real scalars;
the modification to the actual case of QCD, with quarks, gluons and ghosts,
provides no essential difference to what we are going to do. We do need 
initially to include a mass for each field; in the end, all the masses will
be set at their physical values. 

Simply giving masses to gauge fields, even only temporarily,
might appear to be a dangerous procedure. Explicit mass terms
destroy the gauge invariance. If we want to preserve unitarity,
the ghost contribution changes discontinuously when masses are introduced:
the massless limit is no longer
smooth\defref\vDV{H van Dam and M Veltman, Nucl Phys B22 (1970) 397}.
However, we shall assume that the gauge theory has been gauge
fixed first and the only change we shall make is to add masses to all the
fields.  The
mass terms then do not change the total number of degrees of freedom.
It does not matter that, in the presence of the masses, the theory is
not physical; the masses are just a mathematical device. We discuss
this further below.

A simple differentiation with respect to the auxiliary masses gives
$$
{\pd\over\pd m_r^2}\log Z=-\half iZ^{-1}\int \prod _r d\phi _r\left (
\int _C d^4x\,
\phi _r^2(x)\right )\exp\left (iS[\phi ,T]\right )
\eqno(4a)
$$
which is just\ref{\lebellac}
$$
{\pd\over\pd m_r^2}\log Z=-\half i\left < \int _C d^4x\, \phi _r^2(x)\right >
\eqno(4b)
$$
Here, $<\dots >$ denotes a thermal average.
Space-time translation invariance tells us that the thermal average of
$\phi _r^2(x)$ is independent of $x$, and so the $x$ integration is trivial:
$$
{\pd\over\pd m_r^2}\log Z=-{V\over 2T}\left <  \phi _r^2(0)\right >$$$$
=-{V\over 2T}\int {d^4q\over(2\pi)^4} D_r^{12}(q,T)
\eqno(4c)
$$

Here, 
$$
D_r^{12}(q,T)=\left <\int d^4x\,e^{iq.x}\,
\phi _r(0)\phi _r(x)\right >
\eqno(5)
$$
$D_r^{12}(q,T) $ is an element of the familiar $2\times 2$ matrix propagator
${\bf D}_r(q,T)$ of Keldysh-contour\ref{\lebellac}$^,$\defref\keldysh{
L V Keldysh, Sov Phys JETP 20 (1965) 1018
} real-time thermal field theory. This matrix propagator has the general
structure\defref\lvw{N P Landsman and Ch G van Weert,
Physics Reports 145 (1987) 142}\defref\pvlar{
P V Landshoff and A Rebhan, Nucl Phys B410 (1993) 23
}
$$
{\bf D}_r(q,T)={\bf M}(q^0,T)\left (
\matrix{i\hat D_r(q,T)&0\cr
        0&-i\hat D_r^*(q,T)\cr}\right ){\bf M}(q^0,T)
\eqno(6a)
$$
where
$$
\hat D_r(q,T)={1\over q^2-m_r^2-\Pi _r(q,T)}$$$$
{\bf M}(\omega ,T)=\sqrt{n(\omega ,T)}\left (
\matrix{e^{|\omega |/2T} &e^{-\omega /2T}\cr
        e^{\omega /2T}&e^{|\omega |/2T}\cr}\right )
\eqno(6b)
$$
with $n(\omega,T)$ the Bose distribution
$$
n(\omega ,T)={1\over e^{|\omega |/T}-1}
\eqno(6c)
$$
On multiplying out the matrices in (6) we obtain
$$
D_r^{12}(q,T)=-2n(q^0,T)\;e^{|q^0|/2T}e^{-q^0/2T}\,\hbox{Im } \hat D_r(q,t)
\eqno(7)
$$

Insertion of (7) into (4) gives, with (1),
$$
{\pd P\over\pd m_r^2}=-\int {d^4q\over(2\pi)^4} N(q^0,T) \, I_r(q,T)
$$$$N(q^0 ,T)={\epsilon (q^0 )\over e^{q^0 /T}-1}
$$$$I_r(q,T)=
-\hbox{Im }{1\over q^2-m_r^2-\Pi _r(q,T)}
\eqno(8)
$$
Notice that, although we have not written it explicitly, both $\Pi _r(q,T)$
and therefore $I_r(q,T)$ depend on all the masses. 
We assume that the pressure vanishes if all the masses are set to $\infty$,
because then all the particles are static. In order to complete the
calculation of the pressure, we may set all the masses $m_r$ equal to
the same value $m$, and integrate. Then the true pressure, for
all the masses zero, is
$$
P=\sum _r\int _0^{\infty}dm^2\int {d^4q\over(2\pi)^4}N(q^0,T)\,I_r(q,T,m)
\eqno(9)
$$

We have checked explicitly, in the first few orders, 
that (9) does reproduce the diagrams in (1b) with the correct
combinatorial factors. This can be seen by expanding out the
Dyson resummed propagators, upon which the mass integral can
be eliminated by appropriate partial integrations.  This also
works for the more precariously looking case of nonabelian gauge
theories, where Feynman gauge turns out to be the simplest case
to consider. That we are spoiling BRS invariance by giving masses
to all fields does not appear to be a problem. This should not
come as a surprise, since the diagrammatic expansion in (1b)
does not know about BRS. However, we clearly need to have a gauge
fixed theory; otherwise even the interaction-free pressure would
come out wrong.\ref\vDV

We now argue that (9) is not infrared divergent. First, it is evident that
any infrared divergences that may be present in the self energies
$\Pi _r(q,T,m)$ for some values of its arguments will not cause a 
divergence in $P$; the Dyson resummation, which puts $\Pi _r$ into the
denominator of $I_r$, has seen to this. Rather, what might cause trouble is
a zero of the denominator of an $I_r$, or of $N(q^0,T)$.

What is responsible for the increased sensitivity to the infrared regime
at finite temperature is the singularity of 
$N(q^0,T)$ at $q^0=0$. Near $q^0=0$, it behaves as $\epsilon (q^0 )T/q^0$.
However, this is rendered harmless by $I_r$.
$\epsilon(q^0)I_r(q,T)$ is proportional to the spectral density defined by
$$\eqalign{
\rho_r(q,T)&=\left <\int d^4x\,e^{iq.x}\,[\phi _r(x),\phi _r(0)]\right >\cr
           &=D_r^{21}(q,T)-D_r^{12}(q,T)\cr
           &=2\epsilon (q^0)\,I_r(q,T)
\cr}
\eqno(10)
$$
and the spectral density vanishes at $q_0=0$, for
$D_r^{21}(q,T)=e^{q^0/T}D_r^{12}(q,T)$.
In fact, the integral over the singular piece of $N(q^0,T)$ times the
spectral density gives
$$
\int{dq^0\over2\pi} {\rho_r(q,T)\over q^0}
=-{1\over q^2-m^2-\Pi _r(q,T)}\bigg|_{q^0=0}.
\eqno(11)$$
Using this, we can write (9) as
$$
P= \half T \sum _r\int _0^{\infty}dm^2\int {d^3q\over(2\pi)^3}
{1\over {\bf q}^2+m^2+
\Pi _r(q^0=0,{\bf q},T)}
$$$$
+\sum _r\int _0^{\infty}dm^2\int {d^4q\over(2\pi)^4}
\left(N(q^0,T)-{\epsilon(q^0)T\over
q^0}\right)\,I_r(q,T,m)
\eqno(12)
$$
which separates off what would be identified as the zero-mode contribution
in the imaginary-time formalism. 

The zero-mode contribution is clearly harmless in the
infrared, i.e. $|{\bf q}| \to 0$, as long as the number of spatial dimensions
is greater than two. This is true
even for $m=0$ and for the case where $\Pi_r$ does not furnish a mass term.
At $q^0=0$, one would not
expect singularities at real values of $|{\bf q}|$. Nevertheless,
such singularities can appear at some stages of a perturbative evaluation
of $\Pi_r$. In the magnetostatic sector of QCD, these spacelike singularities
have been termed ``Landau ghosts''\defref\chk{J M Cornwall, W-S Hou,
J E King, Phys Lett 153B (1985) 173}. These are generally assumed
to disappear in more accurate calculations of $\Pi_r$, but even so their
presence would not introduce divergences in (12). The corresponding
singularities are smeared out by the additional integration over $m^2$.

Actually, the zero-mode contribution 
itself is ultraviolet divergent, but this divergence 
can be avoided by using dimensional regularization\defref\ae{P Arnold and
O Espinosa, Phys Rev D47 (1993) 3546}; otherwise it has
to be combined with the second term.
Furthermore, there is a logarithmic divergence in the
integral over $m^2$ for large $m^2$, but this is merely an artifact of the
above separation.
In the complete expression (9), 
$I_r$  behaves as $1/m^4$ when $m^2\to\infty$, assuming that
${\rm Im}\,\Pi_r$ does not blow up in this limit.

Consider now the singularities
that might arise in (12)  from a zero in the denominator of an $I_r$. 
Clearly,
the denominator can vanish only if $\Pi _r(q,T,m)$ is
real. This will be the case for $q^0=0$, whereas for nonvanishing $q_0$
it will happen only in some low
order of a perturbative evaluation. Even then, $\Pi_r$ produces a mass gap (the
plasma frequency), which protects us from infrared divergences
also when $m=0$, while $m>0$ will only improve the infrared
behaviour. Moreover, when evaluated through higher orders, $\Pi_r$ will
eventually produce an imaginary part, which removes the singularity
from the real axis.

Since, in the form (9), the pressure is manifestly infrared finite,
this might provide a better starting point for a correspondingly
improved perturbative evaluation of the pressure than standard
perturbation theory, which, in nonabelian gauge
theories, is expected to have an insurmountable
barrier beyond three loop order. In (9), the primary objects to compute
are self energy diagrams. As long as these are kept in the denominator
of (8), all infrared problems should be eliminated. When this becomes
essential, one clearly is going beyond ordinary perturbation theory:
the final result will not have a power-series expansion in the coupling.
Of course, a practical evaluation
of (12) will have to cope with a great number of technical difficulties,
but it does seem to offer a way to circumvent the no-go argument against
a perturbative approach to the pressure. 

\bigskip
{\sl 
This research is supported in part by the EU Programme ``Human Capital
and Mobility", Network ``Physics at High Energy Colliders'', contract
CHRX-CT93-0357 (DG 12 COMA), and by PPARC.}
\vfil\eject
\medskip\immediate\closeout\rfile\writestoppt
\baselineskip=14pt{{\bf References}}\bigskip{\frenchspacing%
\parindent=20pt\escapechar=` \input refs.tmp\bigskip}\nonfrenchspacing
\bye